\documentclass[12pt,a4paper]{article}

\usepackage{amsmath}
\usepackage{amssymb}
\usepackage{epsfig}
\usepackage{graphicx}
\usepackage{cite}

\newcommand{\capdef}{}
\newcommand{\mycaption}[2][\capdef]{\renewcommand{\capdef}{#2}%
       \caption[#1]{{\footnotesize #2}}}
\makeatletter
\renewcommand{\fnum@table}{\textbf{\tablename~\thetable}}
\renewcommand{\fnum@figure}{\textbf{\figurename~\thefigure}}
\makeatother

\newcounter{myenumi}

\renewcommand{\themyenumi}{\roman{myenumi}}
{\end{list}}

\newlength{\myem}
\newcounter{mysubequation}[equation]

\makeatletter
\renewcommand{\section}{\@startsection{section}{1}{0em}{-\baselineskip}%
{\baselineskip}{\normalfont\large\bfseries}}
\renewcommand{\subsection}%
{\@startsection{subsection}{2}{0em}{-0.7\baselineskip}%
{0.7\baselineskip}{\normalfont\bfseries}}
\makeatother

\newcommand{\ie}{{\it i.e.}}

\newcommand{\eg}{{\it e.g.}}

\newcommand{\eq}{Eq.}
\newcommand{\eqs}{Eqs.}

\newcommand{\fig}{Figure}
\newcommand{\Fig}{Figure}

\newcommand{\Ref}{Ref.}
\newcommand{\Refs}{Refs.}
\newcommand{\Sec}{Section}

\newcommand{\App}{Appendix}

\newcommand{\Tab}{Table}

\newcommand{\bi}{\begin{itemize}}
\newcommand{\ei}{\end{itemize}}

\newcommand{\U} {\mbox{$^{235}$U}}
\newcommand{\Ur}{\mbox{$^{238}$U}}
\newcommand{\Pu}{\mbox{$^{239}$Pu}}
\newcommand{\Pl}{\mbox{$^{241}$Pu}}

\begin{document}

\begin{titlepage}

\renewcommand{\thefootnote}{\alph{footnote}}

\begin{flushright}
TUM-HEP-552/04\\

\end{flushright}

\vspace*{1.5cm}

\renewcommand{\thefootnote}{\fnsymbol{footnote}}
\setcounter{footnote}{-1}

{\begin{center}
{\Large\textsf{\textbf{Precision spectroscopy with reactor
anti-neutrinos}}}
\end{center}}
\renewcommand{\thefootnote}{\it\alph{footnote}}

\vspace*{.8cm}

{\begin{center} {{\bf
                Patrick~Huber\footnote[1]{\makebox[1.cm]{Email:}
                \sf phuber@ph.tum.de} and
                Thomas~Schwetz\footnote[2]{\makebox[1.cm]{Email:}
                \sf schwetz@ph.tum.de}
                }}
\end{center}}
{\it
\begin{center}
       Physik--Department, Technische Universit\"at M\"unchen\\
       James--Franck--Strasse, D--85748 Garching, Germany
\end{center}}

\vspace*{0.5cm}

\begin{abstract}
In this work we present an accurate parameterization of the
anti-neutrino flux produced by the isotopes \U, \Pu, and \Pl\ in
nuclear reactors. We determine the coefficients of this
parameterization, as well as their covariance matrix, by performing a
fit to spectra inferred from experimentally measured beta spectra.
Subsequently we show that flux shape uncertainties play only a minor
role in the KamLAND experiment, however, we find that future reactor
neutrino experiments to measure the mixing angle $\theta_{13}$ are
sensitive to the fine details of the reactor neutrino spectra.
Finally, we investigate the possibility to determine the isotopic
composition in nuclear reactors through an anti-neutrino measurement.
We find that with a 3 month exposure of a one ton detector the isotope
fractions and the thermal reactor power can be determined at a few
percent accuracy, which may open the possibility of an application for
safeguard or non-proliferation objectives.
\end{abstract}

\vspace*{.5cm}

\end{titlepage}


\renewcommand{\thefootnote}{\arabic{footnote}}
\setcounter{footnote}{0}


\section{Introduction}

Experiments at nuclear reactors have a long tradition in neutrino
physics. Starting from the experimental discovery of the neutrino at
the legendary Cowan-Reines experiment~\cite{Cowan:1956xc}, many
measurements at nuclear power plants have provided valuable
information about neutrinos. For example, the results of the
G{\"o}sgen~\cite{Zacek:1986cu}, Bugey~\cite{Declais:1995su}, Palo
Verde~\cite{Boehm:2001ik}, and CHOOZ~\cite{Apollonio:2002gd}
experiments have lead to stringent limits on electron anti-neutrino
disappearance. Reactor neutrino experiments have become very prominent
again due to the outstanding results of the KamLAND
experiment~\cite{Eguchi:2002dm,kamland:2004}. For a review on reactor neutrino
experiments, see \Ref~\cite{Bemporad:2001qy}. Recently the possibility
to determine the last unknown lepton mixing angle $\theta_{13}$ by a
reactor neutrino experiment with a near and far detector is actively
investigated (see \Ref~\cite{Anderson:2004pk} and references therein).
Building on the experience gathered in oscillation experiments ideas
of ``applied neutrino physics''
appeared~\cite{Bernstein:2001cz,Nieto:2003wd,Rusov:2004xq}: A detector
close to a nuclear reactor could be used for reactor monitoring, 
either for improving the reliability of operation of power reactors or
as a method to accomplish certain safeguard requirements in the
context of international treaties for arms control and
non-proliferation of weapons of mass destruction.

The standard detection process for reactor neutrinos is inverse beta
decay:
\begin{equation}\label{eq:det}
\bar\nu_e + p \to e^+ + n \,.
\end{equation}
The cross section $\sigma(E_\nu)$ for this process is very well
known~\cite{Vogel:1999zy}, at an accuracy better than 1\%. Per fission
roughly 6 electron anti-neutrinos are produced (see \eg\
\Ref~\cite{Bemporad:2001qy}), with energies peaked around
1~MeV. However, for inverse beta decay only neutrinos with energies
above the threshold of 1.8~MeV are relevant. In nuclear reactors
electron anti-neutrinos in that energy range are produced dominantly
by the beta decay of the fission products from the four isotopes $\ell
=$ \U, \Pu, \Ur, \Pl.\footnote{The next to leading contributions come
from the isotopes $^{240}$Pu and $^{242}$Pu and are of the order 0.1\%
or less~\cite{Bemporad:2001qy}. Further sub-leading effects are the
beta decay of $^{239}$U, $^{239}$Np, $^{237}$U (produced by radiative
neutron capture), and corrections to the spectra from fission
fragments due to neutron absorption by these
fragments~\cite{Kopeikin:1997ve}. These effects are relevant for the
low energy part of the anti-neutrino spectrum $E_\nu \lesssim 2$~MeV
and will be neglected in the following. See also
\Ref~\cite{Kopeikin:2003gu}.} We denote the flux from the isotope
$\ell$ by $\phi_\ell(E_\nu)$ in units of anti-neutrinos per fission
and MeV. In \Tab~\ref{tab:Eell} the total number of $\bar\nu_e$ per
fission above 1.8~MeV is given for \U, \Pu, \Pl, and \Ur.

Accurate information on the anti-neutrino flux from \U, \Pu, and \Pl\
can be obtained by the measurement of the beta spectra from the
exposure of these isotopes to thermal
neutrons~\cite{Schreckenbach:1985ep,Hahn:1989zr,VonFeilitzsch:1982jw}.
Subsequently these beta spectra have to be converted into
anti-neutrino spectra, taking into account the large number of beta
branches involved.  These spectra are in excellent agreement with the
direct observation of the anti-neutrino spectrum at the
Bugey~\cite{Achkar:1996rd} and Rovno~\cite{Kopeikin:1997ve} reactors.
The errors on these fluxes are at the level of a few percent. Since
modern reactor neutrino experiments aim at precisions at this level, a
proper treatment of the flux uncertainties becomes necessary.  For
\Ur, which does only undergo fast neutron fission, no similar
measurements exist, and one has to rely on theoretical
calculations~\cite{Vogel:1981bk,Klapdor:1982sf}.

In absence of neutrino oscillations the number of positron events for
a measurement time $T$ in a given positron energy bin $i$ can be
calculated by
\begin{equation}\label{eq:Ni}
N_i = \frac{n_p T}{4\pi L^2} \sum_\ell N^\mathrm{fis}_\ell \, 
\int \mathrm{d}E_\nu \, \sigma(E_\nu) \, \phi_\ell(E_\nu) \, R_i(E_\nu) \,.
\end{equation}
Here $n_p$ is the number of protons in the detector, $L$ is the
distance between reactor core and detector, and $R_i(E_\nu)$ is the
detector response function for the bin $i$ (including energy
resolution and efficiencies). If the initial composition of the
reactor fuel is known, the number of fissions per second
$N^\mathrm{fis}_\ell$ of each isotope $\ell$ can be calculated
accurately (better than 1\%~\cite{Bemporad:2001qy}) at each burn-up
stage by core simulation codes. The thermal power output $P$ of the
reactor is given by $P = \sum_\ell N^\mathrm{fis}_\ell E_\ell$,
where $E_\ell$ is the energy release per fission for the isotope
$\ell$, see \Tab~\ref{tab:Eell}. Since the errors on $E_\ell$ are less
than 0.5\% we will neglect in the following the uncertainty induced by
them. Defining the relative contribution of the element $\ell$ to the
total power\footnote{The {\it power} fractions $f_\ell$ must not be
confused with the relative {\it fission} contributions
$f_\ell^\mathrm{fis} \equiv N^\mathrm{fis}_\ell / \sum_\ell
N^\mathrm{fis}_\ell$. In this case one would obtain
$N^\mathrm{fis}_\ell = f^\mathrm{fis}_\ell P / \langle E \rangle$,
where the mean energy per fission is given by $\langle E \rangle =
\sum_\ell f^\mathrm{fis}_\ell E_\ell$.}  $f_\ell$, the
$N^\mathrm{fis}_\ell$ can be expressed by $P$ and \eq~(\ref{eq:Ni})
becomes
\begin{equation}\label{eq:Ni2}
N_i = \frac{n_p T}{4\pi L^2} \, P\sum_\ell \frac{f_\ell}{E_\ell} \, 
\int \mathrm{d}E_\nu \, \sigma(E_\nu) \, \phi_\ell(E_\nu) \, R_i(E_\nu) 
\quad\mbox{with}\quad 
f_\ell \equiv \frac{N^\mathrm{fis}_\ell E_\ell}{P} \,.
\end{equation}

\begin{table}[t]
  \centering  
  \begin{tabular}{|l|cc|}
     \hline
     $\ell$ & $N^\nu_\ell$ & $E_\ell$ [MeV] \\ 
     \hline
     \U  & $1.92(1\pm0.019)$ & $201.7\pm0.6$ \\
     \Ur & $2.38(1\pm0.020)$ & $205.0\pm0.9$ \\
     \Pu & $1.45(1\pm0.021)$ & $210.0\pm0.9$ \\
     \Pl & $1.83(1\pm0.019)$ & $212.4\pm1.0$ \\
     \hline
  \end{tabular}
  \mycaption{Total number of $\bar\nu_e$ per fission above 1.8~MeV
  (see \Sec~\ref{sec:errors} for details) and energy release
  per fission (reproduced from \Tab~2 of \Ref~\cite{Apollonio:2002gd})
  for the isotopes relevant in nuclear reactors.}
  \label{tab:Eell}
\end{table}

The aim of the present work is to consider various aspects of reactor
neutrino spectroscopy, with a main emphasis on issues related to the
emitted anti-neutrino flux. First, in \Sec~\ref{sec:parameterization}
we present a phenomenological parameterization for the anti-neutrino
fluxes $\phi_\ell$ based on a polynomial of order 5. We show that in
many situations existing parameterizations~\cite{Vogel:1989iv} are not
accurate enough to describe the reactor neutrino spectrum at the
required level of precision. In addition, in \Sec~\ref{sec:errors} we
give a detailed consideration of the errors associated to the reactor
anti-neutrino spectrum and provide them in a suitable form for
implementation in data analyses. 
In the following we consider the implications of uncertainties of
various quantities appearing in \eq~(\ref{eq:Ni2}) for several
experimental configurations. In \Sec~\ref{sec:kamland} we discuss
the impact of errors on $P$, $f_\ell$, and $\phi_\ell$ for the KamLAND
experiment, whereas in \Sec~\ref{sec:theta13} we discuss the relevance
of our new parameterization of $\phi_\ell$ and its errors for future
reactor experiments to measure $\theta_{13}$. 
In \Sec~\ref{sec:close} we consider the potential of an anti-neutrino
detector close to a reactor: In \Sec~\ref{sec:improving} we discuss
the improvement on the flux uncertainties from a near detector, and in
\Sec~\ref{sec:composition} we investigate the possibility of reactor
monitoring by using the anti-neutrino measurement. In contrast to
previous studies~\cite{Bernstein:2001cz,Nieto:2003wd,Rusov:2004xq} we
employ full spectral information, which allows the determination of
the isotopic content of a reactor, \ie\ the fractions $f_\ell$, as
well as the reactor power $P$ without any external information.
We conclude in \Sec~\ref{sec:conclusions}.


\section{A parameterization for the reactor anti-neutrino flux}
\label{sec:parameterization}

In \Refs~\cite{Schreckenbach:1985ep,Hahn:1989zr} anti-neutrino spectra
from the fission products of \U, \Pu, and \Pl\ are determined by
converting the precisely measured associated beta spectra. In this
section we propose a phenomenological parameterization for the reactor
anti-neutrino flux, based on these measurements. Similar as in
\Ref~\cite{Vogel:1989iv} we parameterize the spectrum of a given
element using a polynomial:
\begin{equation}\label{eq:parametrization}
\phi_\ell(E_\nu) = 
\exp\left( \sum_{k=1}^{K_\ell} a_{k\ell} E_\nu^{k-1} \right) \,.
\end{equation}
The coefficients $a_{k\ell}$ are determined by a fit to the data of
\Refs~\cite{Schreckenbach:1985ep,Hahn:1989zr}. To this aim we minimize
the following $\chi^2$-function:
\begin{equation}
\chi^2 = \sum_{i,j} D_i S^{-1}_{ij} D_j
\quad\mbox{with}\quad
D_i \equiv
\sum_{k=1}^{K_\ell} a_{k\ell} (E_\nu^{(i)})^{k-1} 
 - \ln \phi_\ell^{(i)} \,,
\end{equation}
where $E_\nu^{(i)}$ and $\phi_\ell^{(i)} \equiv
\phi_\ell(E_\nu^{(i)})$ are the values of the neutrino energy and the
corresponding anti-neutrino flux, respectively, provided in the tables
of \Refs~\cite{Schreckenbach:1985ep,Hahn:1989zr} for values of the
neutrino energy $E_\nu$ ranging from 1.5 to 9.5~MeV in steps of
0.25~MeV. Since we are fitting the logarithm of the flux the
covariance matrix $S_{ij}$ contains relative errors of the
$\phi_\ell^{(i)}$. For the diagonal elements $S_{ii}$ we take the
errors as given in the tables of
\Refs~\cite{Schreckenbach:1985ep,Hahn:1989zr} (converted from 90\% CL
to 1$\sigma$ and squared), which contain the statistical error from
the beta spectrum measurement, a systematic error on the overall
calibration, and a systematic error from the conversion from beta
to anti-neutrino spectrum. The off-diagonal elements are obtained from
the error on the absolute calibration, which is taken as fully
correlated: $S_{ij} = \sigma_i^\mathrm{cal} \sigma_j^\mathrm{cal}$ for
$i \neq j$. The errors $\sigma_i^\mathrm{cal}$ are given at two
calibration energies for each isotope in
\Refs~\cite{Schreckenbach:1985ep,Hahn:1989zr,VonFeilitzsch:1982jw},
and we interpolate linearly between these reference points. Note that
this procedure assumes that the systematical errors from the
conversion from beta to anti-neutrino spectrum are completely
uncorrelated between different energies.

\begin{figure}[t!]
   \centering \includegraphics[width=0.98\textwidth]{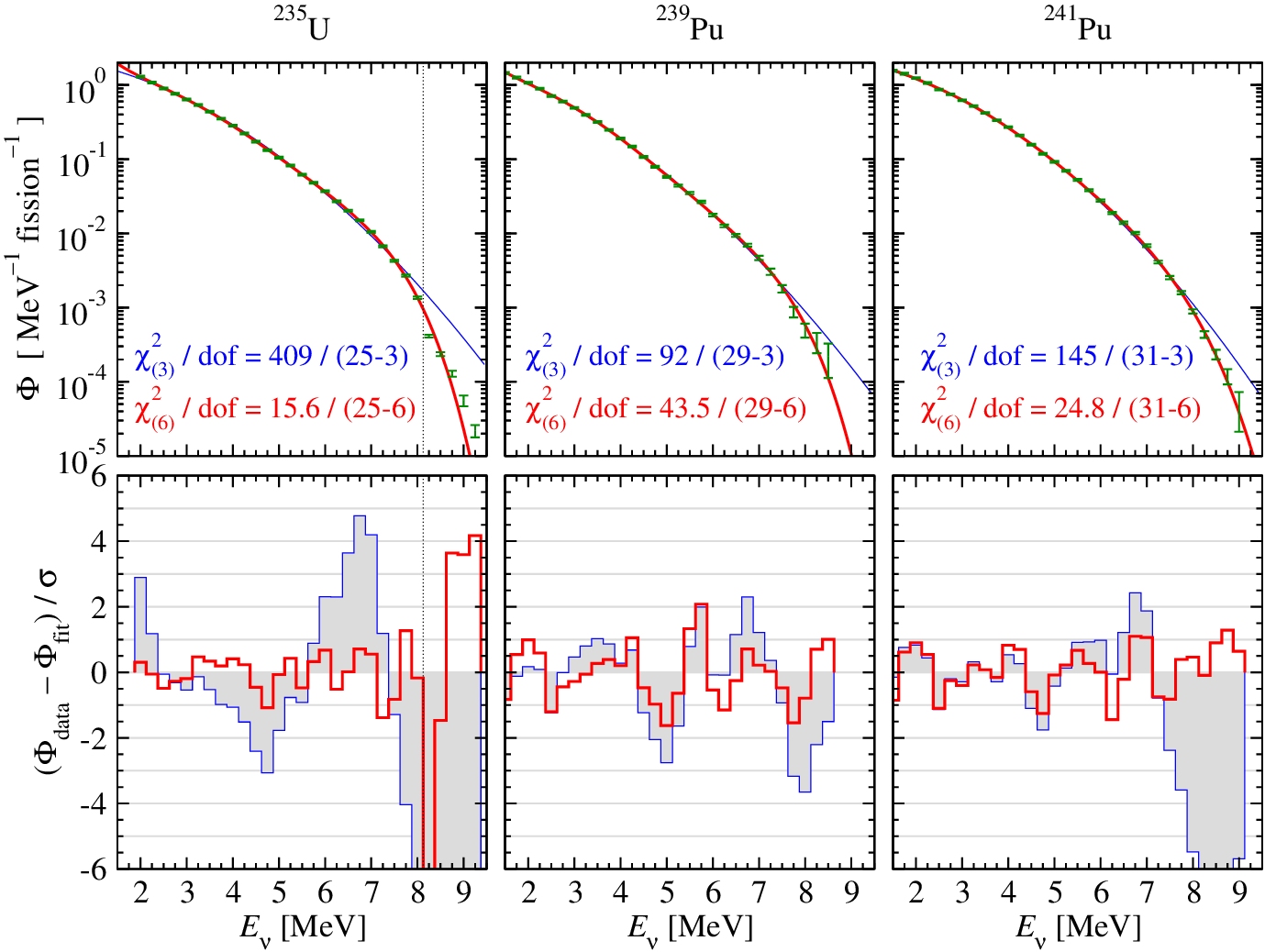}
    \mycaption{(Color online) Illustration of the fit to the data on the
    anti-neutrino spectra from \U~\cite{Schreckenbach:1985ep},
    \Pu~\cite{Hahn:1989zr}, and \Pl~\cite{Hahn:1989zr}. The thick/red
    curves correspond to a 6 parameter fit (polynomial of order 5),
    whereas the thin/blue curves correspond to a 3 parameter fit
    (polynomial of order 2). Also shown are the data with their
    1$\sigma$ error bars and the $\chi^2$ per degree of freedom (number
    of data points minus fitted parameters). In the lower panels we
    show the residuals of the fits. Note that because of correlations
    the shown residuals do not add up to the given $\chi^2$-values.
    The data points to the right of the dotted line in the \U-panels
    are excluded from the fit.} \label{fig:fit} 
\end{figure}

First we have performed a fit of a polynomial of second order ($K_\ell
= 3$). The resulting coefficients $a_{k\ell}$ are given in
\Tab~\ref{tab:three} in \App~\ref{appendix}, and are in reasonable
agreement with the ones obtained in \Ref~\cite{Vogel:1989iv} (some
deviations appear for \Pl). However, we find that the quality of this
three parameter fit is very bad for all three isotopes (the $\chi^2$
is given in \Fig~\ref{fig:fit} in the following). We conclude that at
the level of precision provided by the errors the data cannot be
described with sufficient accuracy by the polynomial of order 2. We
have checked that a reasonable fit is obtained for all three elements
only by going up to a polynomial of order 5, corresponding to $K_\ell =
6$ parameters. The best fit coefficients $a_{k\ell}$ are given in
\Tab~\ref{tab:coefficients} in \App~\ref{appendix}, and the
corresponding anti-neutrino spectra are available in computer readable
format at the web-page \Ref~\cite{webpage}.

Our fit is illustrated in \Fig~\ref{fig:fit}, where we show the
resulting spectra for the 3 and 6 parameter fits in comparison to the
data. Large differences between the 3 and 6 parameter fit are visible
by eye only for the high energy region, where the spectra are very
small and errors are large. However, comparing the corresponding
$\chi^2$-values $\chi^2_{(3)}$ and $\chi^2_{(6)}$ given in the figure
it is obvious that the 6 parameters are necessary to obtain a
reasonable goodness-of-fit. In the lower panels we show the residuals
of the fit, \ie, for each data point $i$ we plot
$(\phi^{(i)}_{\ell\,\mathrm{data}} - \phi^{(i)}_{\ell\,\mathrm{fit}})
/ \sigma_i$, where the error is obtained from the covariance matrix
$S$ by $\sigma_i = \phi^{(i)}_{\ell\,\mathrm{data}} \sqrt{S_{ii}}$.
Note that because of correlations between the
$\phi^{(i)}_{\ell\,\mathrm{data}}$ these residuals do not add up to
the total $\chi^2$. For $\U$ the 3 parameter fit shows rather large
residuals over the full energy range. Since this isotope gives the
main contribution to the reactor anti-neutrino flux it is very
important to model its neutrino spectrum correctly. Let us note that 
we exclude six data points at high neutrino energies from the fit. 
The change in
the spectral shape around 8~MeV~\cite{Schreckenbach:1985ep} cannot be
fitted very well by the polynomial,\footnote{In
\Ref~\cite{Kopeikin:1997ve} a term of order $E_\nu^{10}$ is introduced
to model this sharp falloff.} although by accident the 6 parameter fit
gives a reasonable approximation also in this energy range.  Also for
$\Pl$ the high energy range $E_\nu \gtrsim 7$~MeV is important. In
this case it is possible to obtain a good fit from the polynomial of
order 5 even including the high energy part.


\section{Quantifying the anti-neutrino flux uncertainties}
\label{sec:errors}

In this section we discuss in detail the uncertainties on the reactor
anti-neutrino fluxes. In \Tab~\ref{tab:coefficients} in
\App~\ref{appendix} we show the errors $\delta a_{k\ell}$ on the
coefficients of the polynomial as well as their correlation matrix
$\rho^\ell_{kk'}$, as obtained from the fit to the measured beta
spectra. Hence, the covariance matrix $V^\ell$ for the coefficients can be
obtained by
\begin{equation}\label{eq:covar}
V^\ell_{kk'} = \delta a_{k\ell} \, \delta a_{k'\ell} \, \rho^\ell_{kk'} \,.
\end{equation}

From the table one observes that for a given element the coefficients
are strongly correlated or anti-correlated, since for most elements of
the correlation matrix we obtain $|\rho^\ell_{kk'}| \approx 1$.
Therefore, we perform a rotation in the space of the $a_{k\ell}$, such
that the covariance matrix becomes diagonal.  Let us for each isotope
introduce new coefficients $c_{k\ell}$ by
\begin{equation}\label{eq:rot}
a_{k\ell} = \sum_{k'} \mathcal{O}^\ell_{k'k} \, c_{k'\ell} \,,
\end{equation}
where the orthogonal matrix $\mathcal{O}^\ell$ is defined by
\begin{equation}
\mathcal{O}^{\ell} \, V^\ell \, (\mathcal{O}^\ell)^T
= \mbox{diag}\left[ (\delta c_{k\ell})^2 \right] \,.
\end{equation}
Hence, the $\delta c_{k\ell}$ are the (uncorrelated) errors on the
coefficients $c_{k\ell}$. Using Eqs.~(\ref{eq:parametrization}) and
(\ref{eq:rot}) the anti-neutrino flux for the isotope $\ell$ can be
written as
\begin{equation}
\phi_\ell(E_\nu) = 
\exp\left[ \sum_{k=1}^{K_\ell} c_{k\ell} \, p^\ell_k(E_\nu) \right] \,,
\end{equation}
where $p^\ell_k(E_\nu)$ is a polynomial of $E_\nu$ given by
\begin{equation}\label{eq:poly}
p^\ell_k(E_\nu) = \sum_{k'=1}^{K_\ell} \mathcal{O}^\ell_{kk'}
E_\nu^{k'-1} \,.
\end{equation}

\begin{figure}[t!]
   \centering \includegraphics[width=0.6\textwidth]{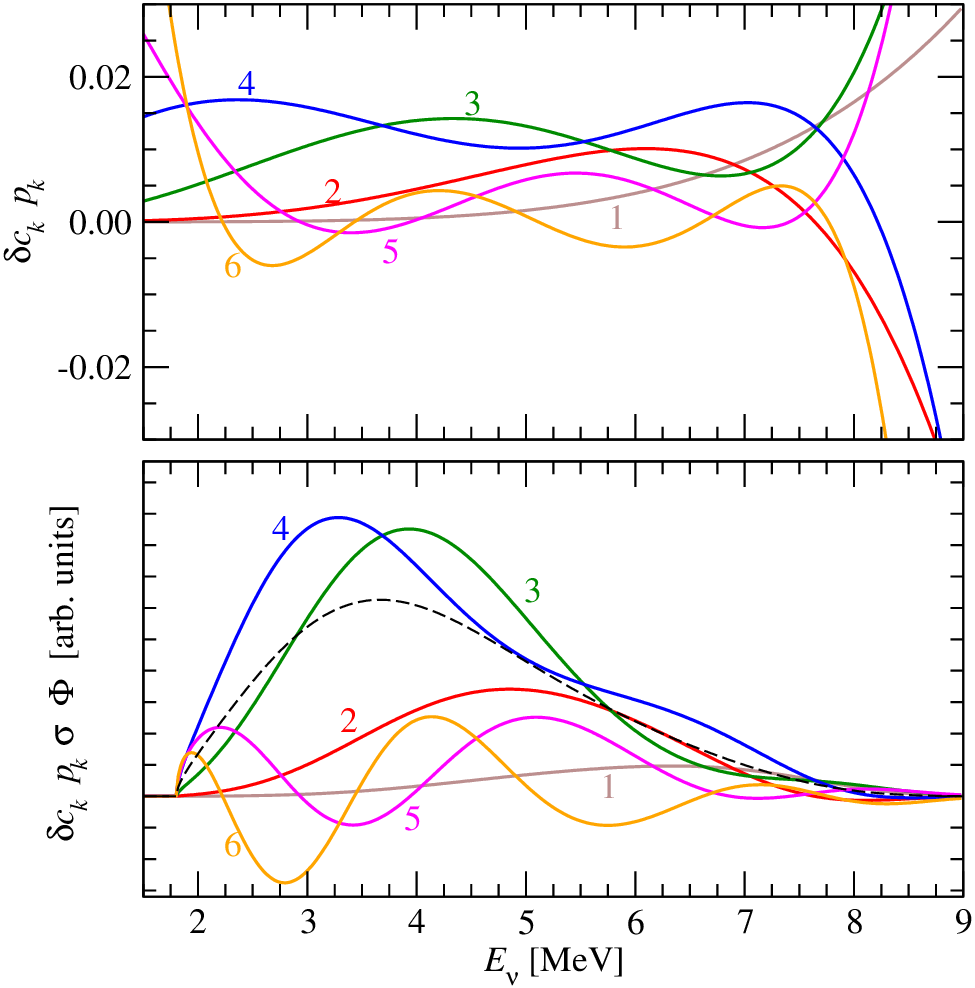}
    \mycaption{(Color online) Uncorrelated anti-neutrino flux
    uncertainties for \U. The upper panel shows the polynomials
    $p^\ell_k(E_\nu)$ given in Eq.~(\ref{eq:poly}) multiplied by the
    corresponding error $\delta c_{k\ell}$. The lower panel shows the
    functions $\delta c_{k\ell} \, p^\ell_k(E_\nu) \, \sigma(E_\nu) \,
    \phi_\ell(E_\nu)$, where $\sigma(E_\nu)$ is the detection cross
    section. The dashed curve corresponds to $\sigma(E_\nu) \,
    \phi_\ell(E_\nu) / 100$.} \label{fig:errors}
\end{figure}

These polynomials describe the uncorrelated contributions to the error
on the anti-neutrino flux. For example, let us consider some
observable $X$, involving the anti-neutrino flux in the following way:
$X = \int \mathrm{d}E_\nu \, h(E_\nu) \, \phi_\ell(E_\nu)$, where
$h(E_\nu)$ is some function of the neutrino energy. Then the error
contribution from the coefficient $c_{k\ell}$ is given by
\begin{equation}\label{eq:example}
\delta X = \delta c_{k\ell} \, \frac{\partial X}{\partial c_{k\ell}} =
\int \mathrm{d}E_\nu \, h(E_\nu) \, \phi_\ell(E_\nu) \, 
\delta c_{k\ell} \,  p^\ell_k(E_\nu) \,.
\end{equation}
Hence, the product $\delta c_{k\ell} \, p^\ell_k(E_\nu)$ is a measure
for the importance of the error $\delta c_{k\ell}$ for any observable.
In the upper panel of \Fig~\ref{fig:errors} we show the polynomials
\eq~(\ref{eq:poly}) weighted by the corresponding error for \U. For
\Pu\ and \Pl\ we obtain very similar results. One observes that in the
relevant range of the anti-neutrino energy the flux uncertainties are
at the level of 2\%. The weighted polynomials $\delta c_{k\ell} \,
p^\ell_k(E_\nu)$ for the isotopes \U, \Pu, \Pl\ are available in
computer readable format at the web-page \Ref~\cite{webpage}. Once
these functions are known, the flux uncertainties on any observable
can be included similar to Eq.~(\ref{eq:example}).

As first simple application let us mention how one can calculate the
number of anti-neutrinos per fission $N^\nu_\ell$ above the threshold
and its uncertainty, as given in \Tab~\ref{tab:Eell}.  Given the best
fit parameters and their covariance matrix for \U, \Pu, \Pl\ we
readily obtain
\begin{equation}\label{eq:Nnutot}
N^\nu_\ell = 
\int_{1.8\,\mathrm{MeV}}^\infty  \phi_\ell (E_\nu) \, \mathrm{d}E_\nu
\,,\qquad
(\delta N^\nu_\ell)^2 = \sum_{kk'} 
\frac{\partial N^\nu_\ell}{\partial a_{k\ell}}
\frac{\partial N^\nu_\ell}{\partial a_{k'\ell}} V^\ell_{kk'} 
=
\sum_k \left(
\frac{\partial N^\nu_\ell}{\partial c_{k\ell}} \,\delta c_{k\ell} 
\right)^2\,.
\end{equation}

In addition to the three isotopes \U, \Pu, \Pl\ also \Ur\ gives a
contribution of a few percent to the reactor anti-neutrino flux. For
this isotope no measurements exist and one has to rely on theoretical
calculations~\cite{Vogel:1981bk,Klapdor:1982sf}. In the following we
will always adopt for \Ur\ the parameterization with the second order
polynomial given in \Ref~\cite{Vogel:1989iv}, which we reproduce in
the last row of \Tab~\ref{tab:three}. In particular, that
parameterization has been used to calculate also the value of
$N^\nu_\ell$ for \Ur\ given in \Tab~\ref{tab:Eell}; the error of 2\%
is an educated guess motivated by the errors obtained for the other
isotopes. Since no covariance matrix of the flux coefficients for \Ur\
is available we will always assume in the following that they are
known exactly. Since the contribution of \Ur\ to the total flux is
rather small, this assumption has very little impact on the conclusions
drawn in this work.


\section{The impact of anti-neutrino flux uncertainties in KamLAND}
\label{sec:kamland}

The KamLAND~\cite{Eguchi:2002dm,kamland:2004} reactor neutrino
experiment is located in the Kamioka mine in Japan and observes the
electron anti-neutrinos emitted by $\sim 16$ nuclear power plants at
distances of $\sim 200$~km. The results of KamLAND have provided
convincing evidence for $\bar\nu_e$ disappearance and are in agreement
with the so-called LMA-MSW solution of the solar neutrino problem
(see, \eg, \Ref~\cite{Maltoni:2002aw}).  The neutrino oscillation
analysis of current KamLAND data is dominated by statistical
errors\footnote{The robustness of the KamLAND results with respect to
statistical fluctuations has been extensively discussed in
\Ref~\cite{Schwetz:2003se}.} and it is and good approximation to
gather various sources of systematical errors into an uncertainty on
the overall number of events. However, in future, if more data are
accumulated statistical errors will decrease and in principle one has
to treat systematical errors more carefully. In this section we
investigate the impact of uncertainties on the anti-neutrino flux for
the determination of oscillation parameters in KamLAND. To this end we
naively extrapolate the size of the data sample published in
\Ref~\cite{Eguchi:2002dm} to a total of five years data taking time by
multiplying the event numbers by the factor $5 \times 356 / 145.1$,
where 145.1 days is the exposure time of the reference sample. For
further details on the KamLAND analysis see
\Refs~\cite{Maltoni:2002aw,Schwetz:2003se}.

In the case of KamLAND one has to generalize the expression for the
number of events per positron energy bin from \eq~(\ref{eq:Ni2}) to
account for the fact that several reactors (labeled by the index $r$)
at different distances $L_r$ contribute to the signal, and that
neutrino oscillations occur:
\begin{equation}\label{eq:NiKL}
N_i = \mathcal{N} \sum_r
\frac{P_r}{L_r^2} \sum_\ell \frac{f_{r\ell}}{E_\ell} \, 
\int dE_\nu \, \sigma(E_\nu) \, \phi_\ell(E_\nu) \, R_i(E_\nu) \,
P_{ee}(L_r / E_\nu) \,.
\end{equation}
Here $\mathcal{N}$ is a normalization constant, $P_r$ and $f_{r\ell}$
are the power output and the element composition of the reactor $r$,
respectively, and $P_{ee}$ is the oscillation probability depending on
the neutrino mass squared difference $\Delta m^2$ and the mixing angle
$\theta$. In our analysis we consider the following contributions to
the covariance matrix $V^\mathrm{KL}$ of the event numbers $N_i$:
\begin{equation}\label{eq:VKL}
V^\mathrm{KL}_{ij} = 
N_i \delta_{ij} + N_i N_j \sigma_\mathrm{det}^2
+ V^\mathrm{flux}_{ij} \,,
\end{equation}
where the first term of the right hand side is the
statistical error and the second term accounts for the over all
normalization error $\sigma_\mathrm{det}$. The uncertainties on the
anti-neutrino flux are included in $V^\mathrm{flux}$, which we take as
\begin{equation}\label{eq:Vflux}
V^\mathrm{flux}_{ij} =
\sum_r 
\frac{\partial N_i}{\partial P_r}
\frac{\partial N_j}{\partial P_r}
(\delta P_r)^2
+
\sum_{r\ell} 
\frac{\partial N_i}{\partial f_{r\ell}}
\frac{\partial N_j}{\partial f_{r\ell}}
(\delta f_{r\ell})^2
+
\sum_{k\ell}
\frac{\partial N_i}{\partial c_{k\ell}}
\frac{\partial N_j}{\partial c_{k\ell}}
(\delta c_{k\ell})^2 \,.
\end{equation}
Here $\delta P_r$ and $\delta f_{r\ell}$ are the errors on the power
output and isotope composition of each reactor, and we assume typical
values of $(\delta P_r)/P_r = 0.02$ and $(\delta f_{r\ell})/f_{r\ell}
= 0.01$. These errors are taken uncorrelated between the reactors.
The last term in \eq~(\ref{eq:Vflux}) takes into account the
uncertainty on the coefficients of the parameterization for the
anti-neutrino fluxes as discussed in Sec.~\ref{sec:errors}.

\begin{figure}[t!]
   \centering \includegraphics[width=0.5\textwidth]{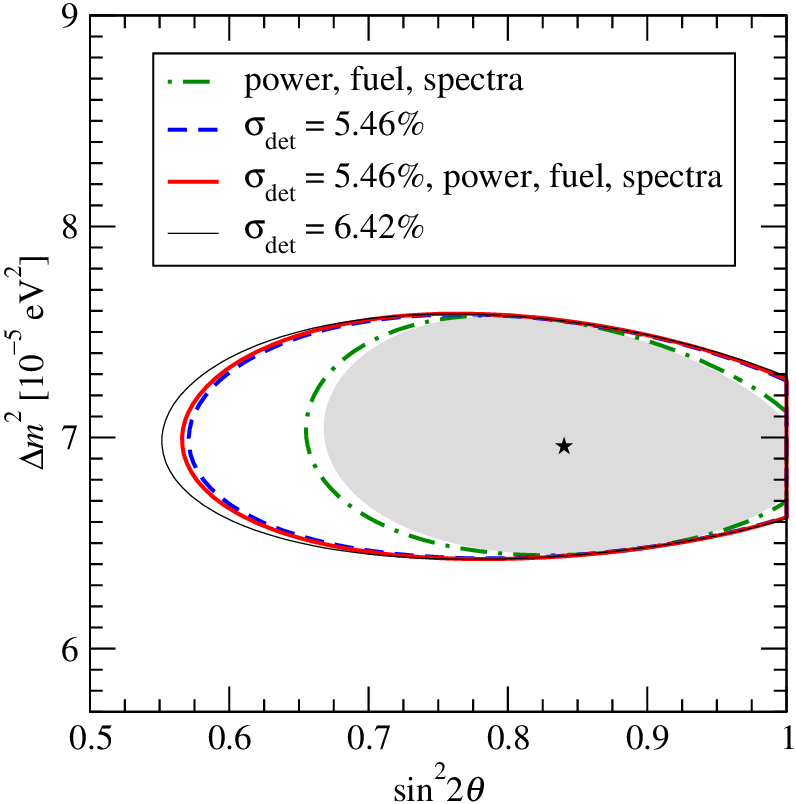}
   \mycaption{(Color online) 3$\sigma$ allowed regions for
   $\sin^22\theta$ and $\Delta m^2$ after 5 years of KamLAND data. The
   shaded region corresponds to statistical errors only, the regions
   delimited by the curves correspond to various assumptions about
   systematical errors. $\sigma_\mathrm{det}$ is a fully correlated
   error on the overall normalization. For the curves labeled ``power,
   fuel, spectra'' we include a 2\% error on the power output of each
   reactor, 1\% error on the fuel composition of each reactor, and the
   uncertainty on the anti-neutrino spectrum as described in the
   text.} \label{fig:kamland} 
\end{figure}

In \Fig~\ref{fig:kamland} we show the 3$\sigma$ allowed regions for
the oscillation parameters after 5 years of KamLAND data for various
assumptions about the systematic errors.  We observe from this figure
that even after 5 years the KamLAND analysis is dominated by the
statistical and the overall normalization errors. The shaded region
corresponds to statistical errors only, for the thin solid curve only
the normalization error of $\sigma_\mathrm{det} =
6.42\%$~\cite{Eguchi:2002dm} is included. In \Tab~2 of
\Ref~\cite{Eguchi:2002dm} various contributions to
$\sigma_\mathrm{det}$ are listed. If the uncertainties related to the
flux are subtracted $\sigma_\mathrm{det}$ is reduced to 5.46\% (see
blue/dashed curve in \Fig~\ref{fig:kamland}). For the red/solid and
green/dash-dotted curves also the flux uncertainties according to
\eq~(\ref{eq:Vflux}) are included. In both cases we find a very small
effect on the oscillation parameters.

To summarize, we find that even for 5 years of KamLAND data the
determination of the oscillation parameters is dominated by
statistical and overall normalization errors. The effect of flux
uncertainties is rather small, and in particular it is not necessary
to fully take into account flux shape errors. The inclusion of only
the normalization errors for the total flux from each element (see
\Tab~\ref{tab:Eell}) leads to nearly identical results as accounting
for the full covariances of the coefficients $a_{k\ell}$. However, the
proper treatment of the flux uncertainties reduces the overall
normalization error. This will become more relevant if in future
KamLAND analyses normalization errors like the uncertainty on the
fiducial volume will be reduced. This will be relevant mainly for the
measurement of the mixing angle, the determination of $\Delta m^2$ is
hardly affected by any of the systematical errors.

Finally, we note that already for the current KamLAND data
sample~\cite{kamland:2004} small differences in the allowed regions
are visible due to the use of our neutrino fluxes, compared to the
parameterization of \Ref~\cite{Vogel:1989iv}. For completeness, we
mention that additional effects like the time evolution of the
individual reactor powers or isotope compositions due to
burn-up~\cite{Murayama:2000iq} may become relevant for future KamLAND
analyses. The investigation of such effects is beyond the scope of the
present work.


\section{Application to future reactor experiments to measure
$\theta_{13}$}
\label{sec:theta13}

Let us now discuss the relevance of the flux uncertainties for reactor
experiments planned to measure the small leptonic mixing angle
$\theta_{13}$. It has been realized that the bound on this angle from
previous experiments~\cite{Apollonio:2002gd,Boehm:2001ik} can be
significantly improved if in addition to a far detector at a distance
of order $2\,\mathrm{km}$ from the reactor a near detector at a few
hundred meters is used. Due to the large number of events, the near
detector provides accurate information on the reactor neutrino flux.
Identical near and far detectors with normalization errors
$\sigma_\mathrm{det}$ below 1\% will provide an accuracy on
$\sin^22\theta_{13}$ of order 0.01 (see, \eg,
\Refs~\cite{Anderson:2004pk,Huber:2003pm,Minakata:2002jv,Martemyanov:2002td,Ardellier:2004ui}).

\begin{figure}[t!]
   \centering \includegraphics[width=0.5\textwidth]{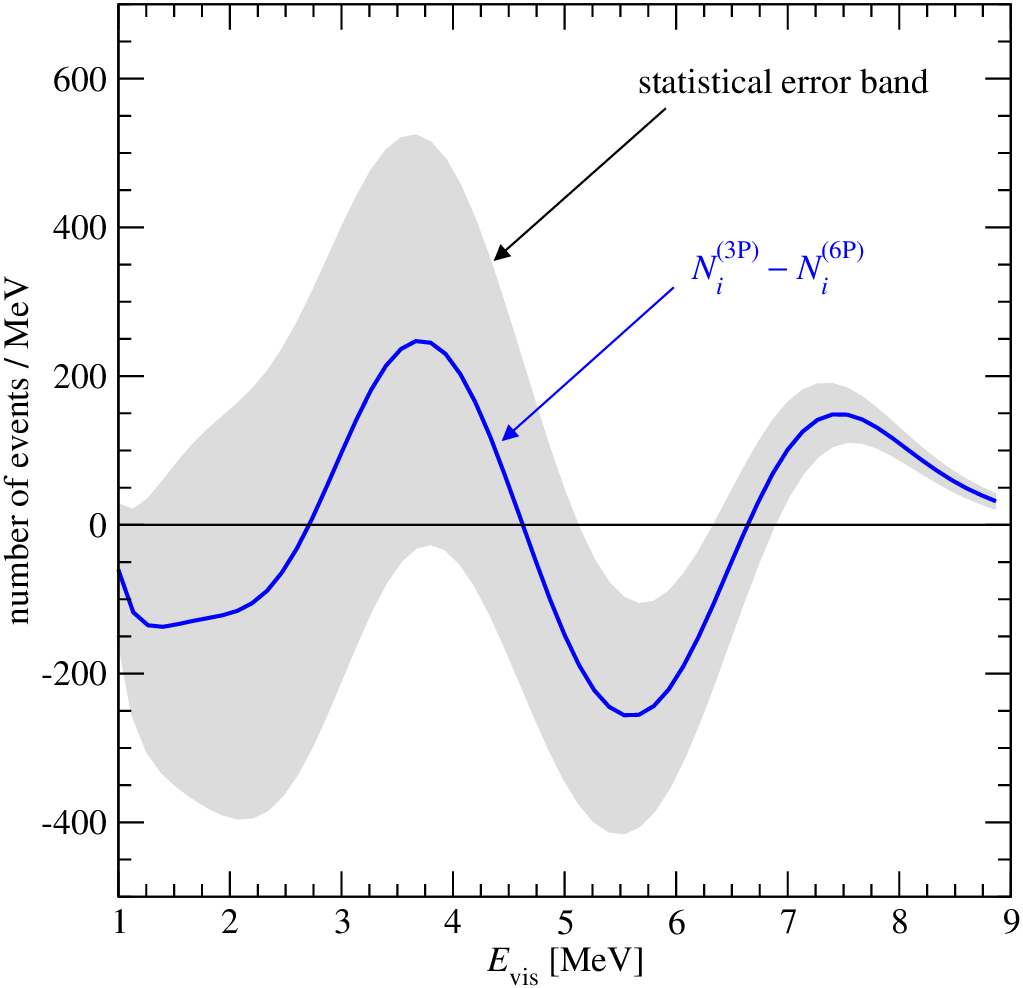}
   \mycaption{(Color online) Difference between the positron spectra
   for no oscillations $N_i^\mathrm{(3P)}$ and $N_i^\mathrm{(6P)}$,
   obtained by using the 3 parameter fit (\Tab~\ref{tab:three}) and 6
   parameter fit (\Tab~\ref{tab:coefficients}) to the anti-neutrino
   spectra, respectively.  We assume a total number of events of 40000
   and a typical isotope composition of $\U : \Pu : \Ur : \Pl = 0.57 :
   0.30 : 0.08 : 0.06$.  The shaded area corresponds to the 1$\sigma$
   statistical error band, \ie, $\pm \sqrt{N_i^\mathrm{(6P)}}$. For 60
   bins in positron energy and statistical errors only we find $\chi^2
   / \mathrm{dof}= 139 / 60$.} \label{fig:3vs6} 
\end{figure}

In Fig~\ref{fig:3vs6} we show the difference between the positron
spectra predicted for no oscillations from the three and six parameter
description of the neutrino spectra. The comparison with the
statistical accuracy for a total number of events of 40000, which is
typical for the far detector of these
experiments~\cite{Anderson:2004pk}, shows that for such experiments a
precise model for the flux is needed. The small differences between
the two parameterizations are clearly distinguishable by the
statistical precision in the far detector: For 60 bins in positron
energy we obtain $\chi^2 / \mathrm{dof}= 139 / 60$, where $\chi^2 =
\sum_i [ N_i^\mathrm{(6P)} - N_i^\mathrm{(3P)}]^2 /
N_i^\mathrm{(6P)}$. Note that in the near detector $\chi^2$ is much
worse, because of the larger number of events.

In the following we adopt the six parameter model for the neutrino fluxes and
investigate the impact of the errors of the coefficients on the sensitivity to
the mixing angle. For simplicity we consider here only two-flavour neutrino
oscillations characterized by the mixing angle $\theta$ and the neutrino
mass-squared difference $\Delta m^2$.  The number of events in a given
positron energy bin $i$ in the detector $A$ ($A=N,F$) can be calculated by
\begin{equation}
N_i^A = 
P\,\frac{\mathcal{N}_A}{L_A^2} \sum_\ell \frac{f_\ell}{E_\ell} \, 
\int dE_\nu \, \sigma(E_\nu) \, \phi_\ell(E_\nu) \, R_i(E_\nu) \,
P_{ee}(L_A / E_\nu) \,.
\end{equation}
Similar to \Ref~\cite{Huber:2003pm} we 
take into account the various systematical errors by writing
\begin{equation}\label{eq:pred}
T_i^A =
(1 + a + b_A) N_i^A + g_A M_i^A +
\sum_\ell 
\zeta_\ell \, f_\ell \, \frac{\partial N_i^A}{\partial f_\ell} +
\sum_{k\ell}
\xi_{k\ell} \, \delta c_{k\ell} \frac{\partial N_i^A}{\partial c_{k\ell}} 
\,.
\end{equation}
Here the parameters $a$ and $b_A$ describe the uncertainty on the
reactor power and the detector normalizations, respectively. The term
$g_A M_i^A$ accounts for the energy calibration (see
\Ref~\cite{Huber:2003pm} for details) and the last two terms in
\eq~(\ref{eq:pred}) describe the uncertainty on the isotope fractions
and the coefficients $c_{k\ell}$, respectively. To test the
oscillation parameters we consider the $\chi^2$
\begin{equation}\label{eq:chisq}
\chi^2 = \sum_{i,A} \frac{ \left( T_i^A - O_i^A \right)^2 }{O_i^A} + 
\left(\frac{a}{\sigma_a} \right)^2 + 
\sum_A \left[
\left(\frac{b_A}{\sigma_\mathrm{det}} \right)^2 +
\left(\frac{g_A}{\sigma_\mathrm{cal}} \right)^2 
\right]+
\sum_\ell \left(\frac{\zeta_\ell}{\sigma_f} \right)^2 +
\sum_{k\ell} \xi_{k\ell}^2  \,,
\end{equation}
where the $T_i^A$ depend on $\theta$ and $\Delta m^2$, whereas $O_i^A =
N_i^A(\theta = \Delta m^2 = 0)$. For each value of $\theta$ and $\Delta m^2$
\eq~(\ref{eq:chisq}) has to be minimized with respect to the ``pulls''
$a, b_A, g_A, \zeta_\ell, \xi_{k\ell}$. 

\begin{figure}[t!]
   \centering \includegraphics[width=0.6\textwidth]{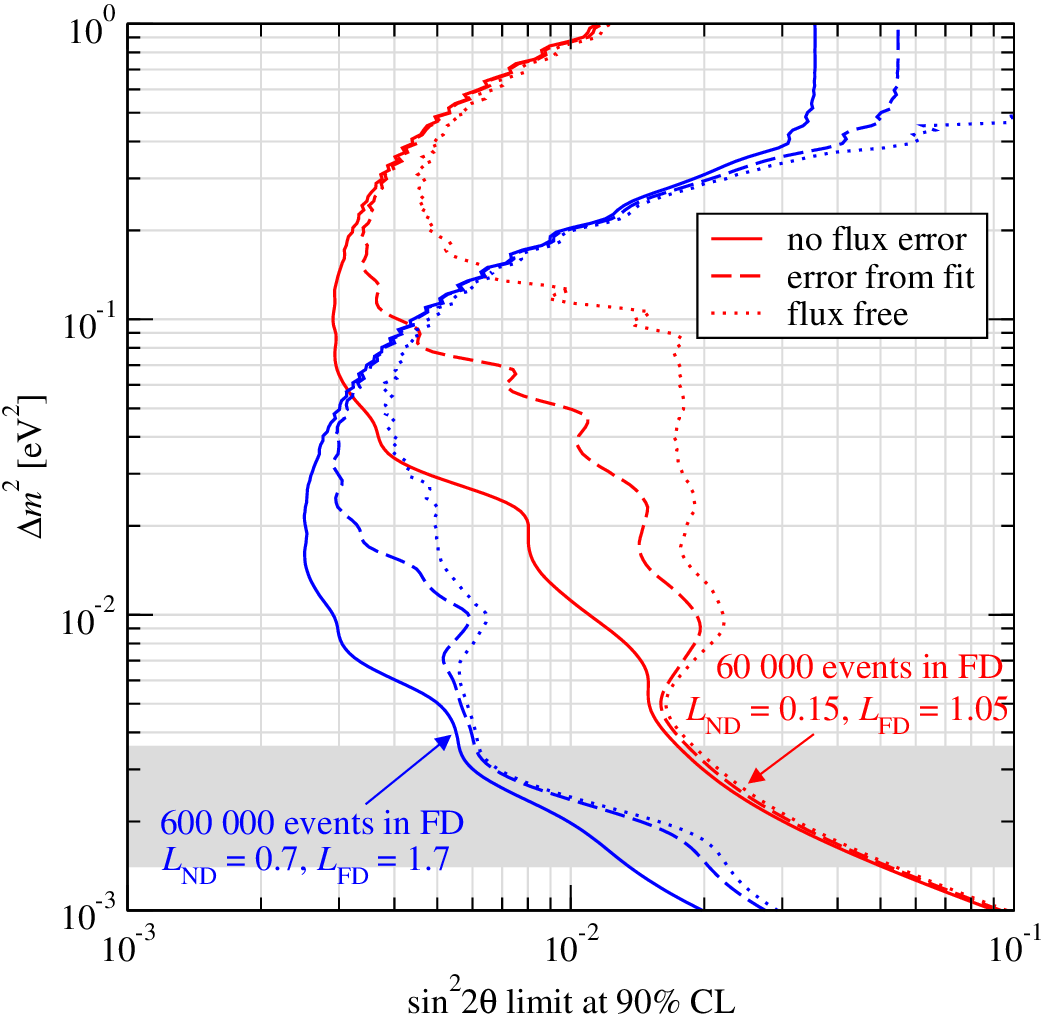}%
   \mycaption{(Color online) The 90\% CL limit on $\sin^22\theta$ from
   reactor neutrino experiments with near and far detectors as a
   function of $\Delta m^2$. The bound is shown for two experimental
   configurations as indicated in the figure. The solid curves
   correspond to no errors on the reactor neutrino flux, for the
   dashed curves the covariance matrix from the fit to the beta
   spectra is used, and for the dotted curves the coefficients for the
   anti-neutrino flux are treated as free parameters in the fit. In
   all cases we have assumed $\sigma_a = 1\%$ for the uncertainty of
   the reactor power, $\sigma_\mathrm{det} = 0.6\%$ for the detector
   normalization, $\sigma_\mathrm{cal} = 0.5\%$ for the energy
   calibration, and $\sigma_f = 1\%$ for the error on the isotope
   fractions. The shaded region indicates the range of $\Delta m^2$
   allowed at 3$\sigma$ from atmospheric neutrino data.}
   \label{fig:th13bound} 
\end{figure}

In \Fig~\ref{fig:th13bound} we show the results of this analysis for a
wide range of $\Delta m^2$. The values of $\Delta m^2$ relevant for
the $\theta_{13}$ measurement are constrained by atmospheric neutrino
data to the interval $1.4 \cdot 10^{-3}\,\mathrm{eV}^2 \le \Delta m^2
\le 3.6 \cdot 10^{-3}\,\mathrm{eV}^2$ at 3$\sigma$ (shaded region in
\Fig~\ref{fig:th13bound}). The regions of $\Delta m^2$ up to
$1\,\mathrm{eV}^2$ could be relevant for oscillations into
hypothetical sterile
neutrinos~\cite{Mikaelyan:1998yg,Kopeikin:2003uu}. The solid curves in
\Fig~\ref{fig:th13bound} are calculated for perfectly known
anti-neutrino flux, \ie, by fixing the coefficients $\xi_{k\ell}$ in
\eqs~(\ref{eq:pred}) and (\ref{eq:chisq}) to zero. For the dashed
curves we use the errors obtained in the fit to the beta spectra as
discussed in Sec.~\ref{sec:errors}. For the dotted curves the
anti-neutrino flux coefficients are treated as free parameters in the
fit, \ie, we drop the last term in \eq~(\ref{eq:chisq}) when
minimizing with respect to $\xi_{k\ell}$.
In \Fig~\ref{fig:th13bound} we consider two experimental
configurations. One corresponds to an experiment with $6\cdot 10^4$
events in the far detector for no oscillations and near and far
detector baselines of $L_\mathrm{ND} = 0.15 \, \mathrm{km}$ and
$L_\mathrm{FD} = 1.05 \, \mathrm{km}$. This set-up is similar to the
Double-Chooz proposal~\cite{Ardellier:2004ui}. For the second
configuration we have assumed a somewhat larger near detector baseline
of $L_\mathrm{ND} = 0.7 \, \mathrm{km}$, a far detector baseline
optimized for $\Delta m^2 \sim 2 \cdot 10^{-3}\,\mathrm{eV}^2$ of
$L_\mathrm{ND} = 1.7 \, \mathrm{km}$, and a rather high luminosity of
$6\cdot 10^5$ events.

Let us first discuss the region of $\Delta m^2$ relevant for the
$\theta_{13}$ measurement. We observe from the figure that in the case
of $L_\mathrm{ND} = 0.15 \, \mathrm{km}$ the limit does hardly depend
on the assumptions concerning the flux uncertainty. Even the flux free
limit is not much worse than the limit for no error on the flux, since
at these short distances the near detector provides a very accurate
determination of the anti-neutrino flux. In contrast, the flux
uncertainty has some impact on the $\theta_{13}$ measurement if the
near detector baseline is somewhat larger. In that case oscillations
start to build up already between reactor and near detector and the
uncertainties on the initial flux become relevant (see also \Fig~12 in
\Ref~\cite{Huber:2003pm}). 
In the region $\Delta m^2 \gtrsim 5 \cdot 10^{-3}\,\mathrm{eV}^2$ the
main information relevant for the limit is provided by the near
detector.  Hence the flux uncertainties become even more relevant for
both configurations in that region. We note that around $\Delta m^2
\sim 2 \, (7) \cdot 10^{-1}\,\mathrm{eV}^2$ for the big (small)
experiment the limit again becomes independent of the flux
uncertainty. In that region rather fast oscillations occur at the near
detector which still can be resolved by the detector, but cannot be
mimicked by adjusting the coefficients of the flux parameterization. 
Obviously, no limit can be obtained for the flux free analysis in the
averaging regime of very high $\Delta m^2$.


\section{Potential of an anti-neutrino detector close to a reactor}
\label{sec:close}

In this section we investigate the potential of an anti-neutrino
detector very close to a reactor, where ``very close'' is defined by
the requirement that neutrino oscillations do not occur. This could be
for example a near detector of the experiments considered in the
previous section, if it is situated close enough to the reactor. In
Subsec.~\ref{sec:improving} we investigate to what extent the
uncertainty on the anti-neutrino flux can be reduced by such a
measurement, whereas in Subsec.~\ref{sec:composition} we consider the
possibility to determine the isotope composition of the reactor core.

To this aim we write the theoretical prediction for the number of
events in bin $i$ given in \eq~(\ref{eq:pred}) as
\begin{equation}
T_i = O_i + \sum_\alpha p_\alpha \Phi_\alpha^i \,,
\end{equation}
where we drop the detector index $A$ and we use the fact that $N_i = O_i$
for no oscillations. The index $\alpha$ runs over all the pulls: 
$p_\alpha = (a, b, g, \zeta_\ell, \xi_{k\ell})$ and the
coefficients $\Phi_\alpha^i$ can be read off from \eq~(\ref{eq:pred}). 
With this notation \eq~(\ref{eq:chisq}) becomes
\begin{equation}\label{eq:chisqpulls}
\chi^2 = \sum_i 
\frac{\left(\sum_\alpha p_\alpha \Phi_\alpha^i \right)^2}{O_i}
+ \sum_\alpha
\left( \frac{p_\alpha}{\delta p_\alpha} \right)^2 \,, 
\end{equation}
where $\delta p_\alpha$ is the error on the pull $p_\alpha$, which can be 
read off from \eq~(\ref{eq:chisq}). Departing from
\eq~(\ref{eq:chisqpulls}) it is straight forward to compute the
improvement of the knowledge on a given parameter $p_\alpha$ due to
the data $O_i$: Because of the quadratic structure of
\eq~(\ref{eq:chisqpulls}) the new covariance matrix $S^\mathrm{new}$
of the $p_\alpha$ can be obtained by inverting
\begin{equation}\label{eq:Snew}
(S^\mathrm{new})^{-1}_{\alpha\beta} =
\frac{1}{2} \, 
\frac{\partial^2 \chi^2}{\partial p_\alpha \partial p_\beta} =
\delta_{\alpha\beta} \, \frac{1}{(\delta p_\alpha)^2} +
\sum_i \frac{\Phi^i_\alpha \Phi^i_\beta}{O_i} \,.
\end{equation}
Note that in general the $p_\alpha$ will be correlated after the
measurement, \ie, $S^\mathrm{new}$ will aquire non-diagonal entries
from the second term in \eq~(\ref{eq:Snew}). The final one sigma
error on a parameter $p_\alpha$ is given by
$\sqrt{S^\mathrm{new}_{\alpha\alpha}}$.

\subsection{Improving our knowledge on the anti-neutrino flux}
\label{sec:improving}

\begin{figure}[t!]
   \centering
   \includegraphics[width=0.6\textwidth]{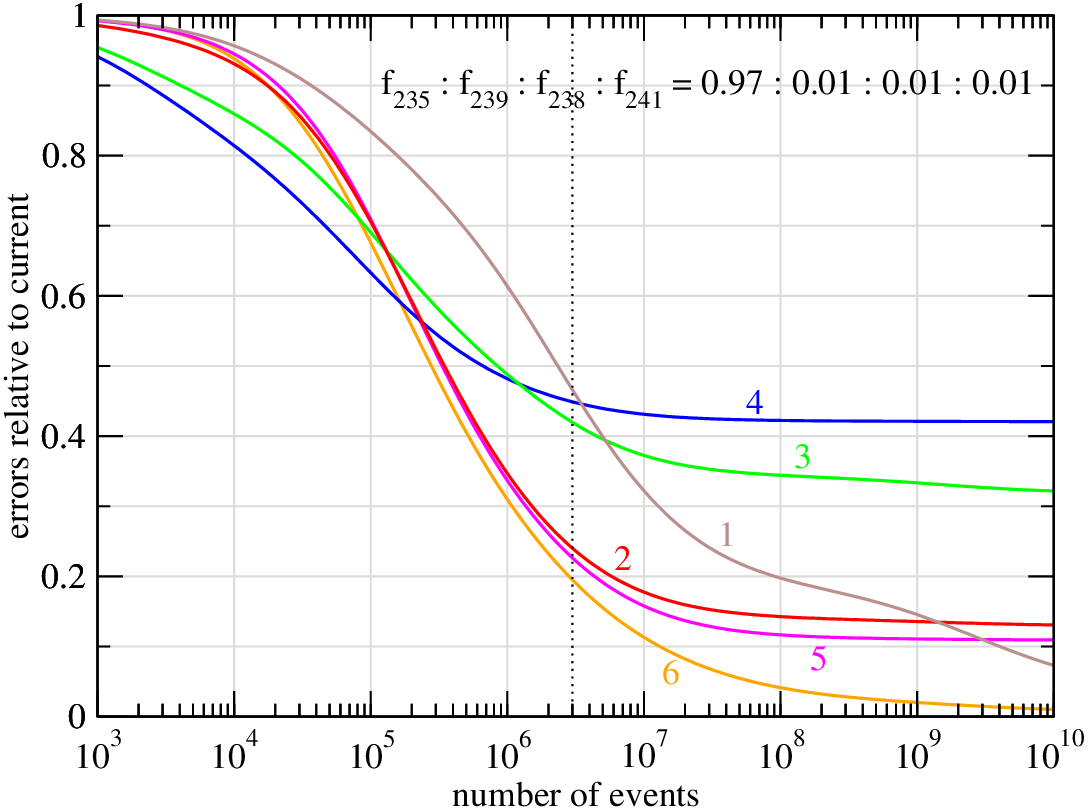}%
   \mycaption{(Color online) Improvement for the anti-neutrino flux
   errors for \U\ as a function of the total number of events in a
   detector close to a reactor. We show
   $\sqrt{S^\mathrm{new}_{\alpha\alpha}}$, where $S^\mathrm{new}$ is
   defined in \eq~(\ref{eq:Snew}) and $\alpha$ runs over the 6 pulls
   associated with the flux-uncertainties shown in
   Fig.~\ref{fig:errors}. We assume that the core contains 97\% $\U$,
   and we take $\sigma_a = 1\%$, $\sigma_\mathrm{det} = 0.6\%$,
   $\sigma_\mathrm{cal} = 0.5\%$, and $\sigma_f = 1\%$. The dotted
   line corresponds roughly to the number of events expected in the
   near detector of the Double-Chooz
   experiment~\cite{Ardellier:2004ui}.} \label{fig:coeff_improvement}
\end{figure}

From \fig~\ref{fig:th13bound} one can see that the limit on the mixing
angle is nearly the same for an analysis with completely free
anti-neutrino flux coefficients (dotted curves) and for the current
errors on them (dashed curves), in the region $\Delta m^2 \lesssim
10^{-2} \, \mathrm{eV}^2$, where oscillations can be neglected in the
near detector. This indicates that the near detector provides a rather
precise determination of the flux on its own. 

This fact is quantified in \fig~\ref{fig:coeff_improvement}, where we
show the improvement of the flux errors for \U\ from the near detector
data with respect to the present errors obtained from the fit to the
beta spectra measurements.  We observe from the figure that for a
number of events $\gtrsim 10^4$ the errors on the flux coefficients
from the anti-neutrino measurement become comparable to the current
errors. To reduce the errors by a factor two roughly $10^7$ events are
needed. To avoid correlations with the flux coefficients from other
isotopes, we assume that the core contains practically only
\U. However, for large number of events the errors on the
coefficients $c_{k\ell}$ become strongly correlated. Especially the
coefficients corresponding to the lines labeled ``2'', ``3'' and
``4'' in \fig~\ref{fig:coeff_improvement} are nearly fully
correlated. This implies that a certain combination of these
coefficients is severely constrained, and one should perform a
diagonalization of the covariance matrix (similar as described in
\Sec~\ref{sec:errors}) to obtain again uncorrelated flux uncertainties
in analogy to \fig~\ref{fig:errors}. Note that the modes labeled ``5''
and ``6'' corresponding to relatively ``fast oscillations'' (compare
\fig~\ref{fig:errors}) can be determined rather good by the
anti-neutrino measurement. We find only modest correlations of these
coefficients.

In general also sizable fractions of \Pu, \Pl\ and \Ur\ will be
present in the reactor. In this case all coefficients will become
correlated. It might be possible to disentangle the contributions of
the various isotopes taking into account precise information on the
time evolution of the reactor composition. The improvement for a given
isotope depends strongly on the relative amount of this isotope in the
core.

\subsection{Determination of the isotope composition of a reactor}
\label{sec:composition}

In this subsection we investigate the possibility to determine the
isotope composition of a reactor core by a nearby anti-neutrino
detector. This could lead to applications of neutrino spectroscopy for
reactor monitoring, either for improving the reliability of operation
of power reactors or as a method to accomplish certain safeguard and
non-proliferation objectives. In both cases the price tag of a
moderately sized detector is small compared to the overall cost or
benefit. Therefore the applicability of neutrino spectroscopy seems to
depend only on its performance compared to existing technologies. For
example the accuracy in the determination of the thermal power of
civil power reactors as used for the production of electricity
typically is in the range $0.6\% - 1.5 \%$~\cite{Apollonio:2002gd}.
The isotopic composition usually is not measured in situ but deduced
from the time development of the reactor thermal power and the initial
isotopic composition by using detailed reactor core simulation tools
and is typically accurate at the percent level~\cite{Nieto:2003wd}. 

On the other hand, one requires for safeguard purposes to achieve a
sensitivity which allows to quickly detect the diversion of weapons
grade material at the level of one critical 
mass~\cite{Bernstein:2001cz}, \eg\ for Plutonium this is approximately
$10\,\mathrm{kg}$. The average power reactor contains three tons of
fissionable material of which roughly $40\%$ have been converted to
Plutonium by the end of the fuel life time, thus $10\,\mathrm{kg}$
Plutonium correspond to $\sim 0.8\%$ of the total content of
Plutonium, which is beyond the accuracy levels for in situ monitoring
today. The determination of the isotopic composition by the
traditional methods furthermore relies on the assumption that the
operator of the reactor collaborates. 
In the following we will show that neutrino spectroscopy has the
potential to reach a sensitivity comparable to existing technologies
and it does \emph{not} require detailed information on the power
history or the initial fuel composition, which is an important
advantage especially in safeguards applications. In contrast to the
existing literature on this
topic~\cite{Bernstein:2001cz,Nieto:2003wd,Rusov:2004xq} we use the
full spectral information and therefore do not require an independent
determination of the reactor power. In general, any safeguard regime
based only on the total rate suffers from two problems: The first one
is related to the availability of reliable information on the thermal
power, whereas the second one is related to the fact that for most
reactor types the diversion of core inventory is only possible during
refueling, \ie\ when the reactor is switched off and there is no
neutrino flux. Thus in order to detect any diversion in this period an
absolute measurement of the neutrino flux as well as of the thermal
power is required. Moreover the composition of the new fuel has to be
known exactly in order to predict the spectrum which is expected in
case of no diversion.

\begin{figure}[t!]
   \centering \includegraphics[width=0.8\textwidth]{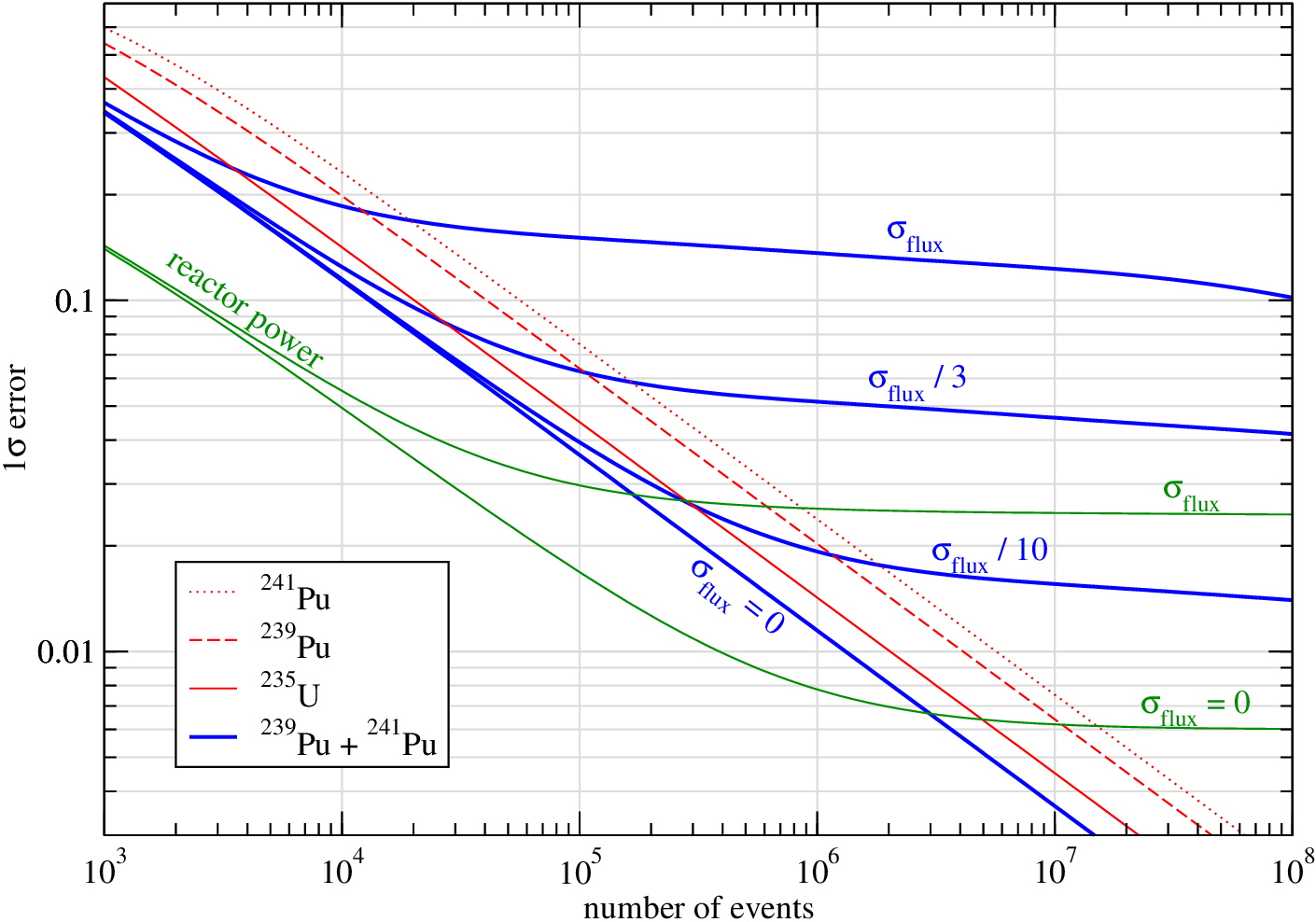}%
   \mycaption{(Color online) The 1$\sigma$ error on the fraction of
   \U, \Pu, and \Pl\ as a function of the number of anti-neutrino
   events. The think curves correspond to the error on the sum of \Pu\
   and \Pl. The straight lines are calculated for perfectly known
   anti-neutrino fluxes. For the sum of \Pu\ and \Pl\ we show also the
   result assuming the present anti-neutrino flux uncertainties
   (``$\sigma_\mathrm{flux}$''), and uncertainties reduced by factors
   three (``$\sigma_\mathrm{flux} / 3$'') and ten
   (``$\sigma_\mathrm{flux} / 10$''). Also shown is the relative
   1$\sigma$ error on the reactor power for $\sigma_\mathrm{flux} = 0$
   and present flux errors. We assume an isotope composition
   $\U:\Pu:\Ur:\Pl = 0.4:0.4:0.1:0.1$ and a detector normalization
   uncertainty $\sigma_\mathrm{det} = 0.6\%$.}
   \label{fig:fraction_lumi}
\end{figure}

In the following we perform a fit where the isotope fractions $f_\ell$ are
treated as free parameters, subject to the condition $\sum f_\ell =
1$. We impose no external information on the reactor power, \ie, no
knowledge at all about the reactor is assumed. This means that we set
$1/(\delta p_\alpha)^2 = 0$ in \eq~(\ref{eq:Snew}) for $\alpha$
corresponding to $\sigma_a$ and $\sigma_f$. The determination of the
isotope fractions and the power is solely based on the differences
between the anti-neutrino spectra emitted by the four isotopes. In
\Fig~\ref{fig:fraction_lumi} we show the 1$\sigma$ accuracy obtained
on the isotope fractions and the reactor power as a function of the
anti-neutrino events. To give an example, for a detector with one ton
fiducial mass at a distance of ten meters from a reactor with three GW
thermal power roughly $10^6$ events are expected within 3 months of
measurement time.

\begin{figure}[t!]
   \centering \includegraphics[width=0.5\textwidth]{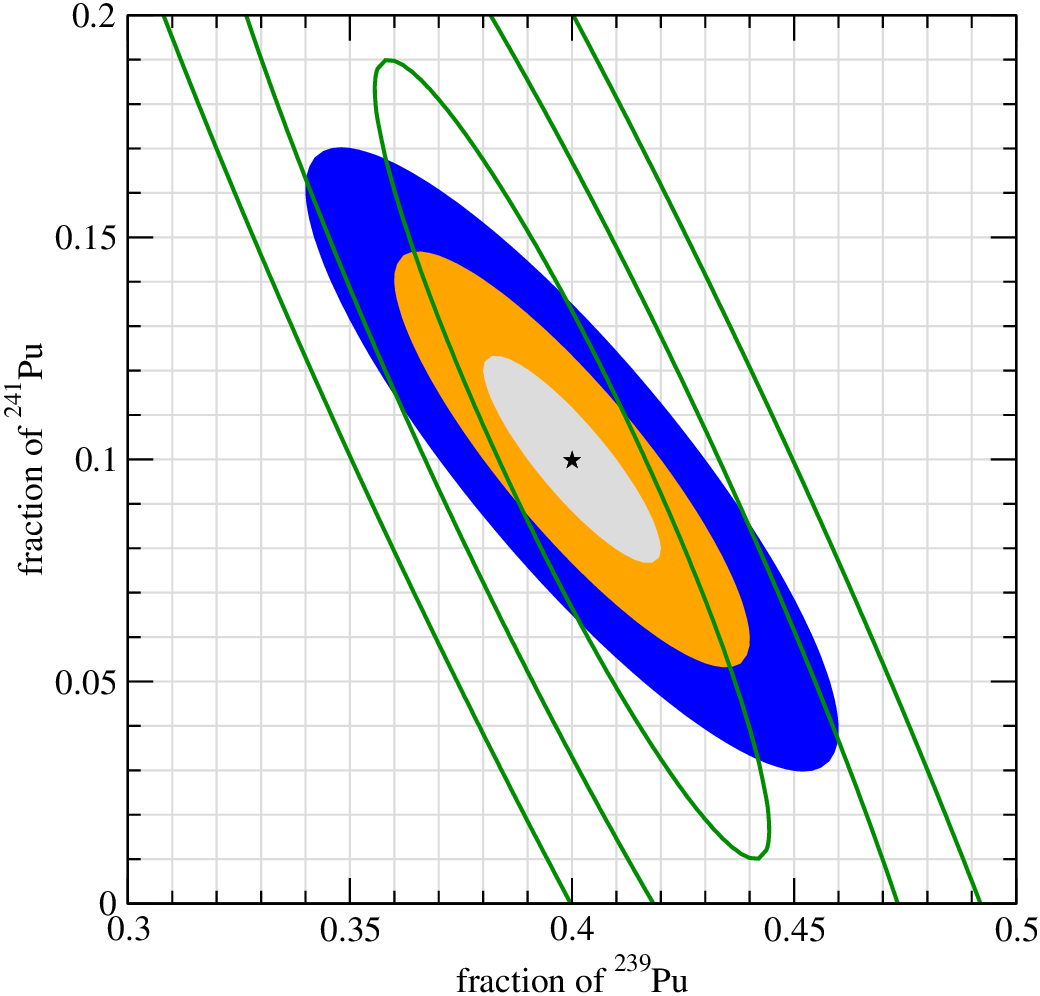}%
   \mycaption{(Color online) Contours of $\Delta\chi^2 = 1,4,9$ in the
   plane of the \Pu\ and \Pl\ isotope fractions for $10^6$
   anti-neutrino events. The colored regions correspond to perfectly
   known anti-neutrino flux shapes, whereas for the curves we assume
   uncertainties on the flux coefficients three times smaller than the
   present errors. We adopt the same isotope composition and
   $\sigma_\mathrm{det}$ as in \Fig~\ref{fig:fraction_lumi}. Reactor
   power and \U\ fraction are treated as a free parameters.}
   \label{fig:PuContours}
\end{figure}

First, one can see from \Fig~\ref{fig:fraction_lumi} that for $\gtrsim
10^5$ events a rather precise determination of the reactor power at
the $\lesssim 3\%$ level is possible, given the current uncertainty on
the anti-neutrino fluxes. For perfectly known fluxes the power
accuracy is limited by the systematical uncertainty of the detector
normalization. Second, for $\gtrsim 10^6$ events also the isotope
fractions of \U, \Pu, and \Pl\ can be determined at the percent level
if no errors on the anti-neutrino fluxes are taken into
account.\footnote{We consider the fractions of \U, \Pu, and \Pl\ as
independent parameters, and determine the \Ur\ fraction by the
constraint $\sum f_\ell = 1$.} The accuracy on the sum of the \Pu\ and
\Pl\ fractions is clearly better than the one on the individual
fractions. This is a consequence of the strong anti-correlation
between the two Pu isotopes, which we illustrate in
\Fig~\ref{fig:PuContours}, where $\chi^2$ contours in the \Pu--\Pl\
plane are shown for $10^6$ events. Note that for safeguard
applications actually the sum of both Plutonium isotopes is the
interesting quantity.

From \Fig~\ref{fig:fraction_lumi} one can see that to determine the
isotope composition a precise knowledge of the emitted fluxes is
necessary. With present errors the 1$\sigma$ accuracy is limited to
$\gtrsim 10\%$. To reach a determination at the percent level the
errors on the coefficients of the flux parameterization have to be
reduced by a factor of three to ten. A factor three would be
approximately achieved by the near detector of an experiment like
Double-Chooz~\cite{Ardellier:2004ui}.

Let us note that in this analysis we do not take into account
additional information such as the time evolution and reactor burn-up,
or information from various traditional safeguard methods. The main
conclusion from the above results is that anti-neutrino spectroscopy
may play an important role for reactor monitoring, especially since
one expects significant synergies due to the combination with
alternative technologies.

\section{Summary and conclusions}
\label{sec:conclusions}

In this work we have presented an accurate parameterization of the
anti-neutrino flux produced by the isotopes \U, \Pu, and \Pl\ in
nuclear reactors. We use a polynomial of order 5 and determine the
coefficients by performing a fit to spectra inferred from
experimentally measured beta spectra. Furthermore, the correlated
errors on these coefficients are determined from the fit.

Subsequently we investigate the impact of the flux uncertainties for 
the KamLAND experiment and future reactor experiments to measure the
mixing angle $\theta_{13}$.  We show that flux shape uncertainties can
be safely neglected in the KamLAND experiment, however the proper
treatment of the errors associated to the anti-neutrino flux reduces
somewhat the overall systematic error in KamLAND, which has some
impact on the determination of the mixing angle. Future high precision
reactor neutrino experiments with a far detector at distances of order
$2\,\mathrm{km}$ and a near detector at hundreds of meters are
sensitive to the fine details of the reactor neutrino spectra. We find
that a parameterization based on a polynomial of order two is not
accurate enough to describe the anti-neutrino spectrum at the required
level of precision. If the near detector is located at distances
$\gtrsim 500$ meters the flux uncertainties are relevant for the
$\theta_{13}$ measurements. Moreover, in searches for sterile
neutrinos at values of $ \Delta m^2\gtrsim 10^{-2}\,\mathrm{eV}^2$ the
main information is provided by the near detector, and hence the
inclusion of anti-neutrino flux uncertainties is essential.

Finally, we have investigated the potential of a detector very close
to a reactor to improve on the uncertainties of the anti-neutrino
fluxes, and to determine the isotopic composition in nuclear reactors
through an anti-neutrino measurement. We find that without any
external knowledge on the reactor a three month exposure of a one ton
detector allows the determination of the isotope fractions and the
thermal reactor power at a few percent accuracy. This may open the
possibility of an application for safeguard or non-proliferation
objectives, which does not rely on information on the reactor thermal
power or on the initial fuel composition, and hence neutrino
spectroscopy can provide information complementary to traditional
monitoring methods. To achieve this goal a reduction of the present
errors on the anti-neutrino fluxes of about a factor of three is
necessary, which naturally can be obtained from the data of the near
detector of a Double-Chooz like experiment.

\subsection*{Acknowledgments}

We thank Michele Maltoni for discussions on the KamLAND analysis, and
Herv{\'e} de Kerret for communication on the CHOOZ experiment. This work has
been supported by the ``Sonderforschungsbereich 375 f{\"u}r
Astro-Teilchenphysik der Deutschen For\-schungs\-ge\-mein\-schaft''.

\begin{appendix}

\section{Results of the fits to the anti-neutrino spectra}
\label{appendix}

In this appendix we give the best-fit coefficients for the polynomials
used to parameterize the anti-neutrino flux of the isotopes \U, \Pu,
and \Pl\ according to \eq~(\ref{eq:parametrization}). In
\Tab~\ref{tab:three} the coefficients for the polynomial of order 2
are given, wheres in \Tab~\ref{tab:coefficients} we display the
coefficients, their errors and the correlation matrix for the
polynomial of order 5.

\begin{table}[h]
  \centering
  \begin{tabular}{|c|ccc|}
  \hline
  $\ell$ & $a_{1\ell}$ & $a_{2\ell}$ & $a_{3\ell}$ \\
  \hline
  \U  & $0.904$ & $-0.184$ & $-0.0878$  \\
  \Pu & $1.162$ & $-0.392$ & $-0.0790$ \\
  \Pl & $0.852$ & $-0.126$ & $-0.1037$ \\
  \hline
  \Ur & $0.976$ & $-0.162$ & $-0.0790$  \\
  \hline
  \end{tabular}
  \mycaption{Coefficients of the polynomial of order 2. For \U, \Pu,
  \Pl\ the numbers are obtained from a fit to the data from
  \Refs~\cite{Schreckenbach:1985ep,Hahn:1989zr}, for \Ur\ we reproduce
  the values given in \Ref~\cite{Vogel:1989iv}. }
  \label{tab:three}
\end{table}

\begin{table}[h]
  \centering
  \begin{tabular}{|r|r|r|rrrrrr|}
  \hline\hline
  \multicolumn{3}{|c|}{$\ell = \U$} & \multicolumn{6}{c|}{correlation matrix $\rho^\ell_{kk'}$}\\
  \hline
  $k$ & $a_{k\ell}$ & $\delta a_{k\ell}$ & 1 & 2 & 3 & 4 & 5 & 6 \\
  \hline
1 & $3.519\cdot10^0$ & $7.26\cdot10^{-1}$  & $1.000$ & $-0.996$ & $0.987$ & $-0.973$ & $0.956$ & $-0.938$\\
2 & $-3.517\cdot10^0$ & $8.81\cdot10^{-1}$  & $-0.996$ & $1.000$ & $-0.997$ & $0.989$ & $-0.976$ & $0.962$\\
3 & $1.595\cdot10^0$ & $4.06\cdot10^{-1}$  & $0.987$ & $-0.997$ & $1.000$ & $-0.997$ & $0.990$ & $-0.980$\\
4 & $-4.171\cdot10^{-1}$ & $8.90\cdot10^{-2}$  & $-0.973$ & $0.989$ & $-0.997$ & $1.000$ & $-0.998$ & $0.992$\\
5 & $5.004\cdot10^{-2}$ & $9.34\cdot10^{-3}$  & $0.956$ & $-0.976$ & $0.990$ & $-0.998$ & $1.000$ & $-0.998$\\
6 & $-2.303\cdot10^{-3}$ & $3.77\cdot10^{-4}$  & $-0.938$ & $0.962$ & $-0.980$ & $0.992$ & $-0.998$ & $1.000$\\  
  \hline\hline  
  \multicolumn{3}{|c|}{$\ell = \Pu$} & \multicolumn{6}{c|}{correlation matrix $\rho^\ell_{kk'}$}\\
  \hline
  $k$ & $a_{k\ell}$ & $\delta a_{k\ell}$ & 1 & 2 & 3 & 4 & 5 & 6 \\
  \hline
1 & $2.560\cdot10^0$ & $4.01\cdot10^{-1}$  & $1.000$ & $-0.993$ & $0.977$ & $-0.954$ & $0.928$ & $-0.899$\\
2 & $-2.654\cdot10^0$ & $5.58\cdot10^{-1}$  & $-0.993$ & $1.000$ & $-0.995$ & $0.982$ & $-0.962$ & $0.938$\\
3 & $1.256\cdot10^0$ & $2.91\cdot10^{-1}$  & $0.977$ & $-0.995$ & $1.000$ & $-0.996$ & $0.984$ & $-0.967$\\
4 & $-3.617\cdot10^{-1}$ & $7.17\cdot10^{-2}$  & $-0.954$ & $0.982$ & $-0.996$ & $1.000$ & $-0.996$ & $0.986$\\
5 & $4.547\cdot10^{-2}$ & $8.37\cdot10^{-3}$  & $0.928$ & $-0.962$ & $0.984$ & $-0.996$ & $1.000$ & $-0.997$\\
6 & $-2.143\cdot10^{-3}$ & $3.73\cdot10^{-4}$  & $-0.899$ & $0.938$ & $-0.967$ & $0.986$ & $-0.997$ & $1.000$\\
  \hline\hline  
  \multicolumn{3}{|c|}{$\ell = \Pl$} & \multicolumn{6}{c|}{correlation matrix $\rho^\ell_{kk'}$}\\
  \hline
  $k$ & $a_{k\ell}$ & $\delta a_{k\ell}$ & 1 & 2 & 3 & 4 & 5 & 6 \\
  \hline
1 & $1.487\cdot10^0$ & $3.23\cdot10^{-1}$  & $1.000$ & $-0.991$ & $0.974$ & $-0.950$ & $0.923$ & $-0.893$\\
2 & $-1.038\cdot10^0$ & $4.31\cdot10^{-1}$  & $-0.991$ & $1.000$ & $-0.994$ & $0.980$ & $-0.960$ & $0.936$\\
3 & $4.130\cdot10^{-1}$ & $2.15\cdot10^{-1}$  & $0.974$ & $-0.994$ & $1.000$ & $-0.995$ & $0.984$ & $-0.966$\\
4 & $-1.423\cdot10^{-1}$ & $5.02\cdot10^{-2}$  & $-0.950$ & $0.980$ & $-0.995$ & $1.000$ & $-0.996$ & $0.986$\\
5 & $1.866\cdot10^{-2}$ & $5.54\cdot10^{-3}$  & $0.923$ & $-0.960$ & $0.984$ & $-0.996$ & $1.000$ & $-0.997$\\
6 & $-9.229\cdot10^{-4}$ & $2.33\cdot10^{-4}$  & $-0.893$ & $0.936$ & $-0.966$ & $0.986$ & $-0.997$ & $1.000$\\
  \hline\hline  
  \end{tabular}
  \mycaption{Coefficients $a_{k\ell}$ of the polynomial of order 5 for
  the anti-neutrino flux from elements $\ell =$ \U, \Pu, and \Pl. In
  the column $\delta a_{k\ell}$ the 1$\sigma$ errors on $a_{k\ell}$
  are given. Furthermore the correlation matrix of the errors is
  shown.} \label{tab:coefficients}
\end{table}

\end{appendix}

\newpage


\end{document}